\documentclass[longauth]{aa}  

\usepackage{graphicx}
\usepackage{gensymb}
\usepackage{txfonts}
\usepackage{xcolor}
\def\baysea{\texttt{BaySeAGal}}

\newcommand{\jp}{J-PAS}
\newcommand{\js}{J-spectra}
\newcommand{\mjp}{miniJPAS}

\def\PyDJ{\texttt{Py2DJPAS}}

\def\magauto{\texttt{MAG\_AUTO}}

\def\rb{$r_\mathrm{SDSS}$}

\begin{document}

   \title{The miniJPAS survey:}

   \subtitle{Dissecting galaxy properties across environments with spatially resolved photometry
   }

   \author{J.E. Rodríguez-Martín.
          \inst{\ref{IAA}}
          \and
          R.M. González Delgado, 
          \inst{\ref{IAA}}  
          \and
          L.A. Díaz-García,
          \inst{\ref{IAA}}             
          \and
          G. Martínez-Solaeche,
          \inst{\ref{IAA}}
          \and
          R. García-Benito,
          \inst{\ref{IAA}}
          A. de Amorim,
          \inst{\ref{UFSC}}
          \and
          J. Thainá-Batista
           \inst{\ref{UFSC}}
          \and
          R. Cid Fernandes
           \inst{\ref{UFSC}}
           \and 
           I. Márquez
           \inst{\ref{IAA}}
           \and
           M. Maturi
           \inst{\ref{UH}}
           \and
          A. Fernández-Soto
         \inst{\ref{UC}, \ref{UV}}
           \and
           R. Abramo
           \inst{\ref{USP-IF}}
           \and
           J. Alcaniz
           \inst{\ref{ON}}
           \and
           N. Benítez
           \and
           S. Bonoli
           \inst{\ref{DIPC}, \ref{IKERBASQUE}}
           \and
           S. Carneiro
           \inst{\ref{ON}}
           \and
           A. J. Cenarro \inst{\ref{CEFCACSIC}}
           \and
           D. Cristóbal-Hornillos \inst{\ref{CEFCA}}
           \and
           R. A. Dupke \inst{\ref{ON}}
           \and
           A. Ederoclite \inst{\ref{CEFCACSIC}}
           \and
           A. Hernán-Caballero \inst{\ref{CEFCACSIC}}
           \and
           C. Hernández-Monteagudo \inst{\ref{IAC}, \ref{ULL}}
           \and
           C. López-Sanjuan \inst{\ref{CEFCACSIC}}
           \and
           A. Marín-Franch \inst{\ref{CEFCACSIC}}
           \and
           C. Mendes de Oliveira \inst{\ref{USP}}
           \and
          M. Moles \inst{\ref{CEFCA},\ref{IAA}}
           \and
           L. Sodré \inst{\ref{USP}}
           \and
           K. Taylor \inst{\ref{instruments4}}
           \and
           J. Varela \inst{\ref{CEFCA}}
           \and
          H. Vázquez Ramió \inst{\ref{CEFCACSIC}}
          }

   \institute{Instituto de Astrofísica de Andalucía (IAA-CSIC), P.O.~Box 3004,
                18080 Granada, Spain\\
              \email{julioeroma@iaa.csic.es, julioeroma@gmail.com }
              \label{IAA}
         \and
             Departamento de F\'{\i}sica, Universidade Federal de Santa Catarina, P.O.~Box 476, 88040-900, Florian\'opolis, SC, Brazil \label{UFSC} 
        \and 
            Instituto de Física de Cantabria (CSIC-UC), Avda. Los Castros s/n, 39005 Santander, Spain \label{UC}
        \and
            Unidad Asociada “Grupo de Astrofísica Extragaláctica y Cosmología”, IFCA-CSIC/Universitat de València, València, Spain \label{UV}
        \and
            University of Heidelberg
            \label{UH}
        \and
            Instituto de F\'isica, Universidade de S\~ao Paulo, Rua do Mat\~ao 1371, CEP 05508-090, S\~ao Paulo, Brazil\label{USP-IF}
        \and
            Observatório Nacional – MCTI (ON), Rua Gal. José Cristino 77, São Cristóvão, 20921-400 Rio de Janeiro, Brazil \label{ON}
        \and
            Donostia International Physics Center (DIPC), Manuel Lardizabal Ibilbidea 4, San Sebastián, Spain \label{DIPC}
        \and
            IKERBASQUE, Basque Foundation for Science, 48013 Bilbao, Spain \label{IKERBASQUE}
        \and
            Instituto de Física, Universidade Federal da Bahia, 40210-340 Salvador, BA, Brazil \label{IF-UFB}
        \and
            Centro de Estudios de Física del Cosmos de Aragón (CEFCA), Unidad Asociada al CSIC Plaza San Juan 1, 44001 Teruel, Spain \label{CEFCACSIC}
        \and
            Centro de Estudios de Física del Cosmos de Aragón (CEFCA), Plaza San Juan 1, 44001 Teruel, Spain \label{CEFCA}   
        \and
            Instituto de Astrofísica de Canarias (IAC), C/ Vía Láctea, S/N, E-38205, San Cristóbal de La Laguna, Tenerife, Spain \label{IAC}
        \and
            Departamento de Astrofísica, Universidad de La Laguna (ULL), Avenida Francisco Sánchez, E-38206, San Cristóbal de La Laguna, Tenerife, Spain \label{ULL}
        \and
        Universidade de São Paulo, Instituto de Astronomia, Geofísica e Ciências Atmosféricas, Rua do Matão 1226, 05508-090 São Paulo, Brazil \label{USP}
        \and
        Instruments4, 4121 Pembury Place, La Canada Flintridge, CA 91011, USA \label{instruments4}
             }

   \date{Received September 15, 1996; accepted March 16, 1997}

 
  \abstract
{The Javalambre-Physics of the Accelerating Universe Astrophysical Survey (\jp) is an ongoing observational programme aiming to map thousands of square degrees in the Northern Hemisphere. By combining 56 narrow band photometric filters with a wide field of view, the survey delivers high-quality, IFU-like data suitable for investigating both the physical properties and the evolution of galaxies on local scales, as well as the influence of their environment. Preceding this, the \mjp\ survey observed a 1~deg$^2$ stripe using the same filter system, serving as a test bench and providing the first scientific results. In this study, we explore the spatially resolved properties of galaxies in \mjp\ and assess the role of environment in their evolution. Our sample comprises 51 galaxies, classified by spectral type (red or blue) and environment (group or field). 
We employ our pipeline, \PyDJ, to download scientific images and catalogues, mask nearby sources, homogenise the images to the same  point spread function (PSF), define galactic regions, and extract their photo-spectra. To analyse the radial profiles of the galaxy properties, we use elliptical annuli in fixed steps of $0.7$~\texttt{R\_EFF}, and apply an inside-out segmentation method to investigate their star formation histories (SFHs). The stellar population properties of these regions are derived using \baysea, a Bayesian parametric code for spectral energy distribution (SED) fitting. Additionally, we employ artificial neural networks to estimate the equivalent widths of the key emission lines: $\mathrm{H}\alpha$, $\mathrm{H}\beta$, $\mathrm{[NII]}$, and $\mathrm{[OIII]}$. We find that stellar population and emission line properties display clear trends in a mass density–colour diagram: redder, denser regions tend to be older, more metal-rich, and exhibit lower star  specific star formation rates (sSFR), indicating a more quiescent state compared with bluer, less mass-dense regions. These latter regions also show stronger emission lines. While red and blue galaxies are distinctly separated in these diagrams, environmental classification does not produce a similarly clear separation. The radial profiles of the stellar population properties of the galaxies are consistent with an inside-out formation scenario, based on its SFH analysis. Red and blue galaxies show distinctly different profiles, but we find no significant influence of environment on these properties. We propose that the absence of a strong environmental effect may be attributed to the relatively low stellar mass of the groups in our sample.}



   \maketitle
%

\section{Introduction}

Surveys using integral field units (IFUs) such as SAURON \citep{SAURON2001}, ATLAS-3D \citep{Cappellari2011}, CALIFA \citep{CALIFA2012}, and MaNGA \citep{MANGA2015} have significantly advanced our understanding of galaxy properties and evolution. Local processes are thought to drive galaxy evolution \citep[see][for a review]{Sanchez2020, Sanchez2021}. Galaxies, whether disk- or bulge-dominated, show gradients in colour, stellar mass density, stellar age, specific star formation rate (sSFR), and metallicity, typically with higher mass densities and older populations in the central regions, supporting an inside-out formation scenario \citep[e.g.][]{Bell2000, Rosa2014, Rosa2015}. However, positive age gradients and rejuvenating galaxies have also been observed \citep[e.g.][]{Goddard2017b, Trayford2016, Tanaka2024}.

In addition to IFU surveys, multiwavelength photometric surveys such as ALHAMBRA \citep{Moles2008}, J-PLUS \citep{Cenarro2019}, and S-PLUS \citep{SPLUS2019} provide complementary data, enabling detailed analysis of stellar populations. Studies using these surveys  have contributed valuable insights into star formation rates (SFRs) and stellar populations in spatially resolved galaxies \citep[see e.g.][]{SanRoman2018,Logrono2019}. These surveys are particularly useful for studying large galaxies that would otherwise be inaccessible due to field of view (FoV) limitations \citep{Rahna2025}.

IFU-like surveys also reveal how galaxy environment affects evolution. Studies of star formation history (SFH) using techniques such as full spectral fitting \citep[e.g.][]{Perez2013, Rosa2016} show that galaxy mergers can trigger rapid increases in sSFR, with active galactic nuclei (AGN) driven quenching occurring inside-out and environmental quenching outside-in \citep{Bluck2020, Lin2019b}. Pre-processing in group environments has been shown to alter galaxy properties, with satellite galaxies experiencing enrichment through gas exchange \citep{Schaefer2019}. Moreover, studies by \cite{Ana2024} highlight differences in stellar populations and mass densities between galaxies in voids and filaments.

The Javalambre-Physics of the Accelerating Universe Astrophysical Survey \citep[J-PAS,][]{Benitez2014} is an ongoing survey designed to cover thousands of square degrees in the Northern Hemisphere from the Observatorio Astrofísico de Javalambre \citep[OAJ,][]{Cenarro2014} using the 2.5 m JST/T250 telescope. Its photometric filter system comprises 54 narrow, equally spaced every 100~\AA, with a full width at half maximum (FWHM) of $\sim 145$~\AA, plus two medium band filters, $u_\mathrm{JAVA}$ and J1007, achieving a spectral resolution of $R \sim 60$, comparable to very low-resolution spectroscopy, over the entire optical wavelength range. The camera's effective FoV is $4.2$~$\mathrm{deg}^2$, and it employs JPCam \citep[JPCam,][]{Taylor2014,MarinFranch2017}, a 1.2~Gpixel camera with a pixel scale of $0.23$~arcsec~$\mathrm{pixel}^{-1}$, using 14 CCDs simultaneously. Prior to this, the miniJPAS survey \citep{Bonoli2020} covered 1deg$^2$ along the AEGIS stripe, using the same filter system and providing the first scientific results. This survey was observed using the  J-PAS Pathfinder  (JPAS-PF) camera equipped with a single CCD, the same J-PAS filter set, and with a low-noise detector of one $9.2k \times 9.2k$ pixel. The resulting FoV is 0.27~$\mathrm{deg}^2$ with a pixel scale of $0.23$~arcsec~$\mathrm{pixel}^{-1}$. 

The data from both miniJPAS and J-PAS are ideal for conducting IFU-like studies. Their large FoV allows for the observation of large galaxies within a single pointing, while the filter system enables precise determination of stellar population properties \citep[see][]{Rosa2021,Luis2024}, as well as the estimation of equivalent widths (EWs) for key emission lines \citep{Gines2021,Gines2022} and the detection of quasars using machine learning techniques \citep{Queiroz2023, Rodrigues2023,Gines2023}. Furthermore, the survey's design facilitates the detection of clusters and groups without pre-selection bias in the observed targets \citep[see e.g.][]{Doubrawa2023, Maturi2023}. This is ensured by the survey’s large field of view, which enables the inclusion of all galaxies brighter than the detection limit within the observed area, regardless of their prior classification or known membership. Consequently, the sample is not biased by targeted observations or by pre-selection based on environmental indicators such as colour, morphology, or redshift.  In this work, we aim to explore the spatially resolved properties of galaxies in miniJPAS, taking advantage of the capabilities of the J-PAS filter system and the large FoV to investigate the properties of these galaxies and the role of their environment in their evolution. We also demonstrate the potential of future J-PAS data for similar studies.

This paper is structured as follows: in Sect.~\ref{sec:data} we describe the nature of the \mjp \ data, as well as the criteria used to select our sample, in Sect.~\ref{sec:method} we describe the methodology used for this paper, in Sect.~\ref{sec:results} we study the properties of the spatially resolved galaxies in \mjp \ and the role of the environment in their evolution, and in Sect.~\ref{sec:discuss} we discuss our results. All the magnitudes used in this paper are given in the AB magnitude system \citep{Oke1983}. Throughout this paper, we assume a Lambda cold dark matter ($\Lambda$CDM) cosmology with $h = 0.674$, $\Omega _{\mathrm{M}} = 0.315$, $\Omega _\Lambda = 0.685$, based on the latest results by the \cite{Planck2020}.

\section{Data} \label{sec:data}
\begin{figure*}
    \includegraphics[width=\textwidth]{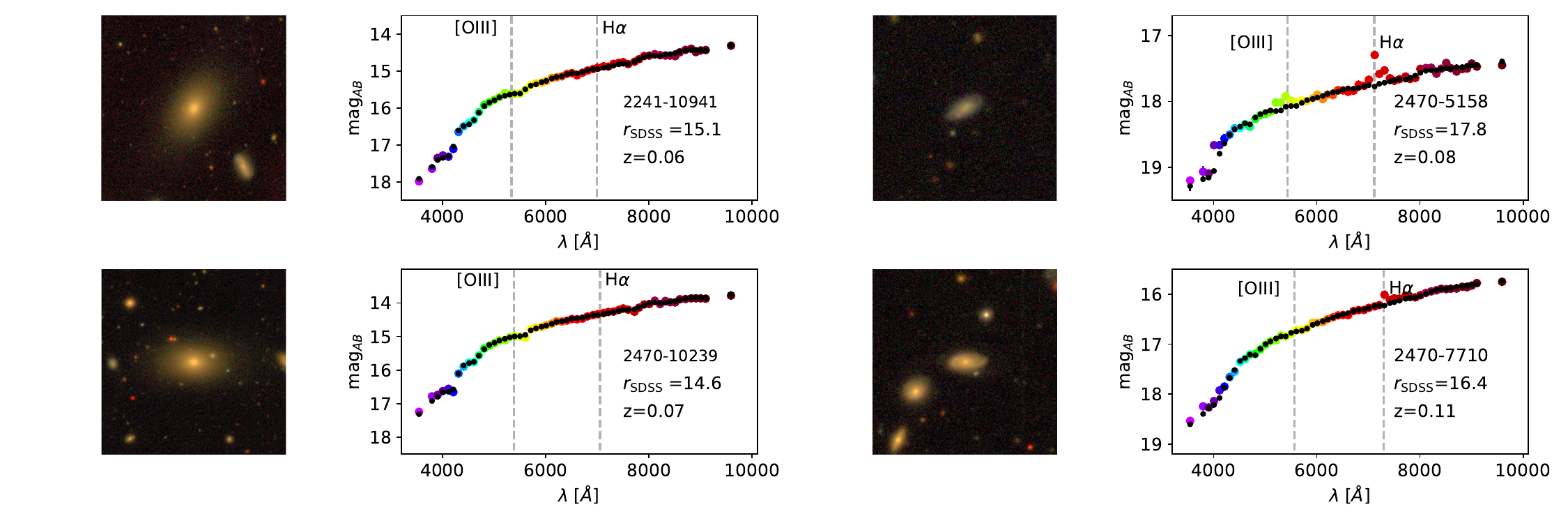}
    \caption{Example of galaxies in the sample and their \js. Left panels show the RGB images of two red galaxies. Second column shows the \magauto \ \js \ of the red galaxies. Third column shows the RGB images of two blue galaxies. Right panels show the \magauto \ \js \ of blue galaxies. Colour points represent the observed magnitudes. Black dots represent the result of the SED-fitting. Grey dashed lines represent the wavelengths at which [OIII] and H$\alpha$ are found. Galaxy spectra are shown in the observer frame, and they correspond to the galaxy in the centre of the image. Redshifts shown correspond to the most likely value (\texttt{PHOTOZ}) of the redshift probability density function \citep[$z$PDF ,see][for further details]{HC2021}.} 
    \label{fig:gal_example}
\end{figure*}

The data used in this study come entirely from the \mjp\ survey \citep{Bonoli2020}. In this section, we describe the selection criteria applied to the galaxies included in our sample (Sect.~\ref{sec:data:select}), and their classification into field and group environments (Sect.~\ref{sec:data:env}). The observational and integrated properties of the galaxies in the sample are summarised in Appendix ~\ref{app:properties}.

\subsection{Sample selection}\label{sec:data:select}

Galaxies have been selected based on the following criteria:  
(i) they are at least twice as large as the FWHM of the worst point spread function (PSF), allowing for at least two extractions ($\mathrm{R\_EFF} > 2''$);  
(ii) they are not edge-on galaxies (ellipticity must be smaller than $0.6$);  
(iii) they exhibit no artifacts or nearby sources that could bias the photometry (\texttt{MASK FLAGS} = 0 for all bands, and \texttt{FLAGS} must not include the flag 1);  
(iv) the objects are classified as galaxies (\texttt{CLASS\_STAR} < 0.1).

With these criteria, we obtain a total of 51 galaxies. These galaxies are subsequently divided into red and blue populations using the selection criterion from \cite{Luis2024}, which is an adaptation of the method proposed by \cite{Luis2019} and previously employed by \cite{Rosa2021} to separate galaxy populations in \mjp\ into red and blue categories. We classify galaxies as red if they satisfy the following condition:
\begin{equation}
    (u-r)_{\mathrm{int}} > 0.16 \left(\log( M_{\star}) - 10\right) - 0.254 \times (z - 0.1) + 1.689,
\end{equation}
where $(u-r)_{\mathrm{int}}$ is the intrinsic $(u-r)$ colour of the galaxy, corrected from extinction, and $M_{\star}$ is the total stellar mass of the galaxy, both calculated from SED-fitting. Otherwise, the galaxy is classified as blue. According to this criterion, we identify 27 blue galaxies and 24 red galaxies. In Fig.~\ref{fig:gal_example} we show some red and blue galaxies, where the differences between the spectral types, such as the different 4000-break amplitudes or the emission lines found in the blue galaxies can be appreciated. The median error of the sample in the \magauto \ photometry are $0.026$~mag, and $0.007$~mag for the fits. Consequently, the error bars shown in Fig.~\ref{fig:gal_example} are typically smaller than the plotted symbols. The general properties of these galaxies can be found in Appendix~\ref{app:properties}.

\subsection{Environmental classification}\label{sec:data:env}

To study the effects of environment, we further split our sample galaxies into field and group galaxies. For this purpose, we use the adaptation of the AMICO code \citep[][]{Maturi2005, AMICO} developed by \cite{Maturi2023} for the \mjp\ data release. This code is based on the Optimal Filtering technique \citep[see][for more details]{Maturi2005, Bellagamba2011}, and provides, among other parameters, a probabilistic association of each galaxy with every detected group or cluster.

A key advantage of this approach is that it operates directly on the full, flux-limited galaxy catalogue, without relying on prior knowledge of known clusters or on targeted observations. This ensures that both cluster and field galaxies are identified from the same homogeneous dataset, thereby minimising pre-selection biases related colour, or morphology. 

This probability has been used in previous studies investigate environmental effects on galaxy evolution \citep{Rosa2022, Julio2022}. Following the same classification as \cite{Rosa2022}, we define galaxies as group members if their highest probabilistic association exceeds $0.8$, and as field galaxies if this value is below $0.1$. These thresholds ensure high-purity samples while limiting cross-contamination between environments. Based on this criterion, we identify 15 red field galaxies, 21 blue field galaxies, 9 red group galaxies, and 6 blue group galaxies. Following the classification scheme used by other works such as \cite{Yang2007,Yang2009,Bluck2020}, where the central galaxy is defined as the most massive galaxy within a dark matter halo. Similarly, we can use the list of Brightest Group Galaxies (BGGs) provided by \cite{Maturi2023}, considering these as the central galaxies of their respective groups. According to this definition, we find a total of 8 galaxies that are the BGGs in their groups. However, given the already limited sample size, we chose not to incorporate this classification in our analysis, as it would reduce the number of galaxies in each category even further. This will be taken into account in future work.

\section{Methodology} \label{sec:method}
In this section we summarise the methods and codes used for the analysis of our data. Our methodology is mainly based on three codes: \PyDJ  \ \citep{Julio2025}, a python tool used to automatise the steps required to obtain the photometric spectra (\js) of the regions of galaxies (Sect.~\ref{sec:codeSP}), \baysea \citep{Rosa2021}, the  code for spectral energy distribution (SED) fitting used to estimate the properties of the regions (Sect.~\ref{sec:SEDfit}), and the Artificial Neural Networks (ANNs) trained by \cite{Gines2021}, used to estimate the EWs of the main emission lines (Sect.~\ref{sec:ANN}).

\subsection{\PyDJ} \label{sec:codeSP}

\PyDJ\  is our custom code developed to streamline and automate the full process required to study spatially resolved galaxies in \jp. In brief, the code begins by downloading the scientific images and tables necessary for the analysis, including key parameters such as galaxy coordinates, zero points, and photometric redshifts (photo-$z$). It then masks nearby sources using both the default masks provided in the \jp\ database and additional custom masks that we generate, including deblended sources.

We adopt $\texttt{NPIX} = 10$ and $rms = 5$, where \texttt{NPIX} is the minimum number of pixels for a detection to be considered valid, and $rms$ defines the threshold level for pixel detection. These values strike a good balance between effectively masking nearby contaminants and preserving the genuine background structure \citep[see][]{Julio2025}.

Next, all images are convolved to match the FWHM of the worst PSF using a Gaussian approximation. This step avoids introducing artificial colour gradients in the photometry \citep[e.g.,][]{Michard2002, Cyprian2010, Molino2014, Liao2023}, and is commonly applied in similar analyses \citep[see e.g.][]{Bertin2002, Darnell2009, Desai2012, SanRoman2018}.

In the final step, the galaxy regions are defined, and the \js\ is computed for each.
We implement three segmentation strategies: (i) Maximum-resolution segmentation: Elliptical rings with a major axis equal to the FWHM of the worst PSF. Rings are added until the median signal-to-noise ratio (S/N) drops below 10. (ii) Fixed-step segmentation: Elliptical rings with steps of $0.7 \times$ \texttt{R\_EFF}, which is the size of the FWHM of the worst PSF compared with the respective galaxy in the sample. (iii) Inside-out segmentation: two regions are defined, an inner region ($a \leq 0.75$~\texttt{R\_EFF}) and an outer region ($0.75 < a \leq 2$~\texttt{R\_EFF}), where $a$ is the semi-major axis. This method is used to analyse SFHs (see Sect.~\ref{sec:SRSFH}). Only regions with a median signal-to-noise ratio (S/N) greater than 5 in filters with pivot wavelengths $\lambda_{\text{pivot}} < 5000$ \AA\ are included in the analysis. This criterion is motivated by the strong correlation between the 4000 \AA\ break and the stellar population properties of galaxies \citep[see e.g.][]{Rosa2005}. Our results show that, for each filter, apertures meeting this S/N threshold yield residuals well below a relative difference of 10~\%, leading to more accurate fits. In contrast, when measurements in a given filter fall below this threshold, residuals increase substantially, often exceeding 10~\%, and there is a systematic underestimation of the observed flux \citep[see][]{Julio2025}. Hence, this restriction is applied. The median number of regions per galaxy produced by the maximum-resolution segmentation is six, both before and after applying the S/N criterion. For the fixed-step segmentation, the corresponding median is six regions per galaxy prior to the S/N cut, and five regions per galaxy once the cut is applied. The median error of the measurements is $0.07$~mag for the homogeneous ring segmentation, and $0.06$~mag for the maximum resolution segmentation. The median error of the fits are $0.015$~mag for both segmentations. After applying the S/N threshold, the median errors  decrease to $0.06$~mag for the measurements and $0.013$~mag for the fits, again for both segmentations.

\subsection{J-spectra fits} \label{sec:SEDfit}
To obtain the SED fitting for the regions of the galaxies, we utilize \baysea \ \citep{Rosa2021}, a parametric SED fitting code that employs a Markov Chain Monte Carlo (MCMC) approach to explore the parameter space. This method generates a sample of parameters that approximates the probability density function (PDF) of the model. In this study, we adopt the same $\tau$-delayed SFH model used in \cite{Rosa2021, Rosa2022} and \cite{Julio2022}.

We retrieve the following parameters: $t_0$ (the time when star formation commenced, in look-back years), $\tau$ (the timescale over which star formation is spread), stellar extinction ($A_V$), stellar metallicity ($Z$), stellar mass ($M_\star$), and we also calculate the mass-weighted and light-weighted ages, rest-frame colours, and star formation rates (SFR).  We retrieve the parameters $t_0$, $\tau$, $Z$, and $A_V$ by marginalisation, taking advantage of the Bayesian framework. To account for extinction, the model spectra are attenuated by a factor $e^{-q_\lambda \tau_V}$, where $\tau_V$ is the dust attenuation parameter in the $V$ band and $q_\lambda \equiv \tau_\lambda/\tau_V$ represents the reddening law. In this work, we adopt the attenuation law of \cite{calzetti2000}, which assumes a single foreground screen with a fixed value of $R_V = 3.1$. We use an uniform prior, with a range of values to look for the solution $t_0 = [0,0.99]$ (in units of the age of the universe, taken from z), $\tau = [0.1, 10]$~Gyr, $A_V = [0,2]$~mag.

In Fig.~\ref{fig:gal_example} we show the quality of the fittings obtained with this code. The models used for the fitting are the version of the \cite{C&B2003} stellar population synthesis models updated by \cite{Plat2019}. As these models do not include emission lines, the fit is constrained by the stellar continuum, leading to the expected differences between observed and modelled magnitudes in the filters affected by line emission, as seen in Fig.~\ref{fig:gal_example}. For this reason, we rely on artificial neural networks (ANNs) to estimate the equivalent widths of the relevant emission lines.

\subsection{ANN for emission lines}\label{sec:ANN}

For estimating the emission line properties, we utilize the ANN trained and applied by \cite{Gines2021,Gines2022}. The ANN used here was specifically trained to predict the EWs of $\mathrm{H}\alpha$, $\mathrm{H}\beta$, $\mathrm{[NII]}$, and $\mathrm{[OIII]}$. The input layer of the network takes the colours of all the filters relative to the filter where $\mathrm{H}\alpha$ is observed. To account for redshift, separate ANNs are trained for each redshift value, ranging from $z=0$ to $z=0.35$ with a step size of $0.001$. The hidden layers optimize the loss function, and the output layer provides estimates for the EWs of $\mathrm{H}\alpha$, $\mathrm{H}\beta$, $\mathrm{[NII]}$, and $\mathrm{[OIII]}$.

The ANN model is trained by means of spectra from the MaNGA and CALIFA surveys, convolved with the filters of \mjp\ to simulate synthetic \js. The spectra span various regions with different characteristics, including star-forming, quiescent, and AGN regions, ensuring that the ANN is trained across a broad range of galaxy types and conditions. This diversity improves the accuracy of the ANN predictions for different galaxy types or regions under study. For a detailed description of error handling and missing data, we refer the reader to the original work by \citep{Gines2021}. The results of that work indicate that the EWs of $\mathrm{H}\alpha$, $\mathrm{H}\beta$, $\mathrm{[NII]}$, and $\mathrm{[OIII]}$ in SDSS galaxies can be estimated with a relative standard deviation of $8.4$~\%, $13.7$~\%, $14.8$~\%, and $15.7$~\%, respectively, and relative biases of $0.03$~\%, $5.0$~\%, $4.8$~\%, and $-6.4$~\%, respectively. In addition, the $\mathrm{[NII]}$/$\mathrm{H}\alpha$ ratio is constrained within 0.092~dex with a bias of $-0.02$~dex, while the $\mathrm{[OIII]}$/$\mathrm{H}\beta$ ratio shows no bias and a dispersion of $0.078$~dex in SDSS galaxies. Furthermore, this methodology was also employed by \cite{Gines2022} in order to retrieve the cosmic evolution of the SFR density, which was found to be in agreement with previous measurements based on the $\mathrm{H}\alpha$ emission line.

\section{Results} \label{sec:results}

This section presents the spatially resolved stellar population properties (Sect \ref{sec:res:srprop}), radial profiles of stellar population parameters  (Sect \ref{sec:radprof}), emission line profiles (Sect \ref{sec:res:EL}), and a comparison between the SFHs of inner and outer regions (Sect \ref{sec:SRSFH}). To streamline the discussion, we define here the key properties and units used throughout this section. For clarity, units will not be repeated in the text. The main properties analyzed are: (i) stellar surface mass density, $\mu_\star$, measured in $M_\odot\mathrm{pc}^{-2}$; we generally refer to its logarithmic value (ii) mass-weighted stellar age, $\left<\log \mathrm{age}\right>_M$, expressed in years (logarithmic scale) (iii) mass-weighted stellar metallicity, $\left<\log(\mathrm{Z}/\mathrm{Z}\odot)\right>$ (iv) visual extinction, $A_V$, given in magnitudes (AB system) (v) sSFR, in units of $\mathrm{Gyr}^{-1}$; we typically refer to its logarithm.

\subsection{Spatially resolved properties}\label{sec:res:srprop}
\begin{figure}
    \centering
    \includegraphics[width=0.4\textwidth]{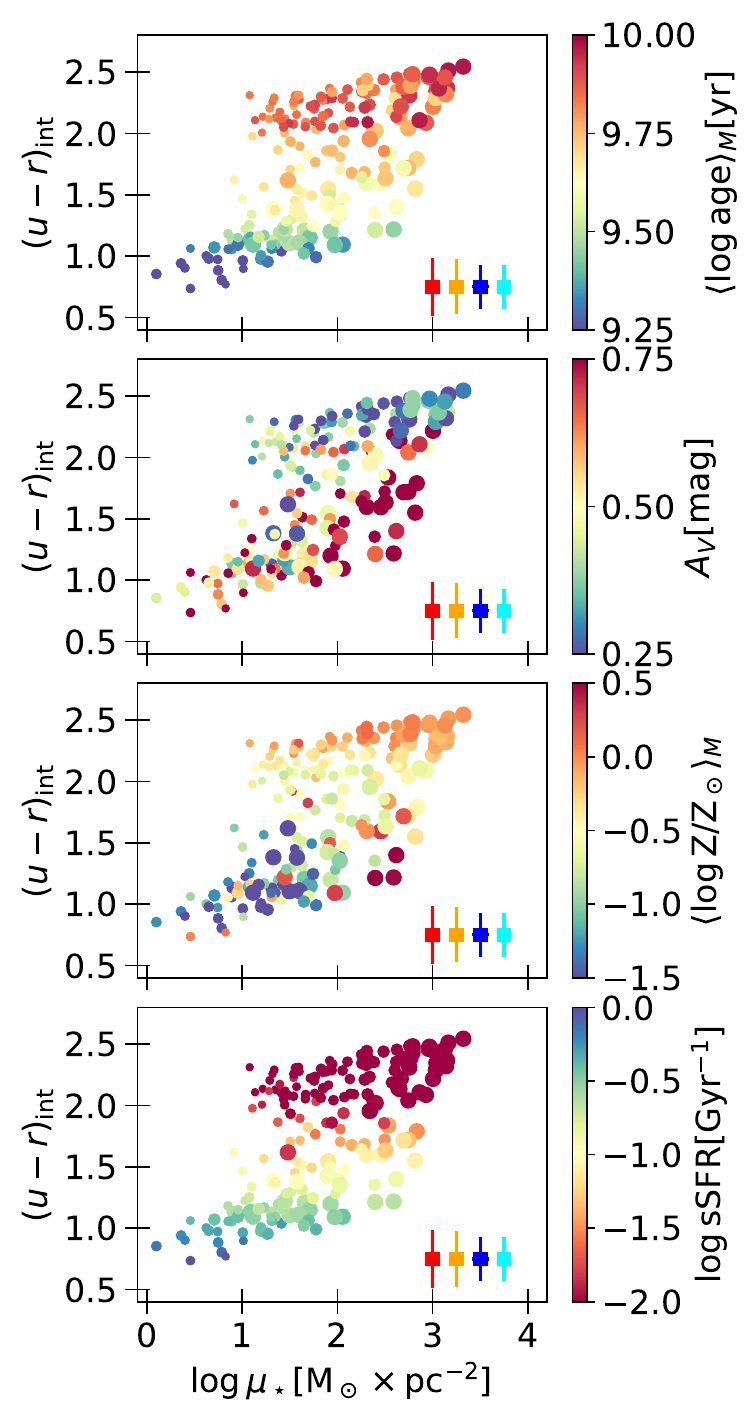}
    \caption[Colour--mass density diagram colour-coded by the main stellar population properties]{Colour--mass density diagram colour coded by the main stellar population properties. Each point represents a region. From  top to bottom: mass-weighted age, extinction, stellar metallicity, and sSFR. Each point represents a different aperture. Squares represent the typical error along each axis colour for each type of galaxy: red galaxies in the field (red), red galaxies in groups (orange), blue galaxies in the field (blue) and blue galaxies in groups (cyan). Point size is inversely proportional to the distance to the galactic centre.}
    \label{fig:colourmassdenSPP}
\end{figure}

We begin our analysis by introducing a diagram analogous to the stellar mass–colour diagram, corrected for extinction \citep{Luis2019} substituting stellar mass with stellar mass surface density. To ensure a balanced representation of data points across the diagram, we apply a fixed-step segmentation method (see Fig.~\ref{fig:colourmassdenSPP}). This approach yields a well-defined distribution of stellar population properties. To aid the interpretation of these diagrams, we note that the median uncertainties of the displayed properties are as follows: $0.15$~dex for the mass-weighted age, 0.19~mag for the extinction $A_V$, $0.39$~dex for the stellar metallicity, and $0.11$~dex for the sSFR. 

Older regions are found in the redder and higher surface mass density parts of the diagram, while younger regions appear bluer, particularly in areas of lower stellar surface mass density. Interestingly, the highest dust extinctions are observed in the blue, high stellar surface mass density regions. On the other hand, the lowest metallicities are found in the blue, low stellar surface mass density areas, whereas some of the most metal-rich regions correspond to blue but dense regions. These points are also associated with regions of higher extinction. A dust-metallicity degeneracy can be taking place, although the values for these points remain below the maximum allowed during the SED fitting process. Furthermore, there is no corresponding increase in the age of these regions, which could have been an alternative parameter for the code to use in reddening the fitted spectra. However, the primary source of uncertainty in these considerations stems from the significant uncertainty in the metallicity values. Overall, redder and denser regions tend to be more metal-rich. These results are in agreement with the radial profiles found in the literature \citep[see e.g.][]{Rosa2014,Rosa2015,Rosa2016,SanRoman2018,Bluck2020,Parikh2021,Abdurro2023}, since the stellar mass surface density decreases steeply with the radial distance. In this regard, galaxy regions with a high stellar mass surface density can be expected to be closer to the centre. It is precisely in the innermost regions where  the oldest and more metal rich stellar populations, particularly for redder galaxies, are usually found. This also aligns with results shown in our diagrams.

Regions that are both red and dense exhibit the lowest sSFR, suggesting they are the most quiescent, in agreement with the radial profiles of surface mass density and sSFR reported by \citet{Rosa2015,Rosa2016} and \cite{Bluck2020}. Broadly speaking, these diagrams reproduce the same relationships observed in stellar mass-colour diagrams based on galaxies’ integrated properties \citep[e.g.,][]{Luis2019,Rosa2021,Rosa2022,Julio2022}, implying once the colour, extinction, and stellar mass surface density of a given region in a galaxy are known, its other stellar population properties can be reliably constrained.

Although both colour and mass density correlate with stellar population properties, colour appears to be the stronger predictor, given the mostly vertical gradient in the diagrams, particularly for the stellar age and the sSFR. This aligns with the results of \citet{Luis2019}, who, using $\tau$-delayed SFHs from \citet{Madau1998} to model galaxies in the ALHAMBRA survey, found that galaxy properties correlate most strongly with colour, but also with size \citep{Luis2019c}. 
Our findings support this, demonstrating that properties such as stellar age and specific star formation rate (sSFR) are linked to the intrinsic $(u - r)_\mathrm{int}$ colour. However, stellar surface mass density also plays a significant role, particularly when distinguishing between blue and red regions.

\begin{figure}
    \centering
    \includegraphics[width=0.45\textwidth]{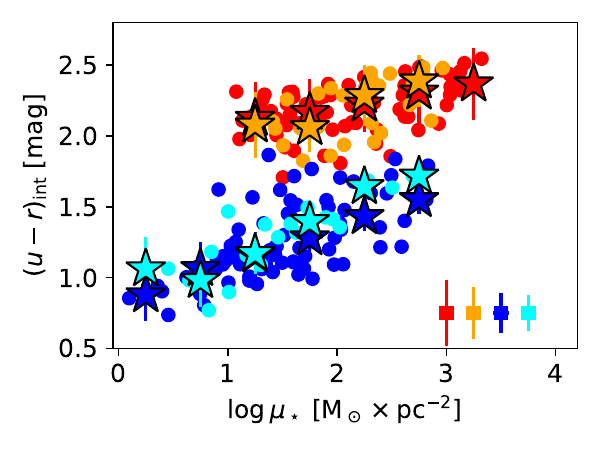}
    \caption[Colour--mass density diagram coloured by the environment and colour of the galaxies]{Colour--mass density diagram coloured by the by the environment and colour of the galaxies. Red points represent regions belonging to the red galaxies in the field, orange is used for the regions of the red galaxies in groups, blue points are for the regions of blue galaxies in the field and cyan points are for regions of blue galaxies in groups.  Stars represent the median value for each galaxy type in each mass density bin. Squares represent the same as in Fig.~\ref{fig:colourmassdenSPP}.}
    \label{fig:colourmassdencolourandenv}
\end{figure}

We also use this diagram to explore the distribution of galaxy regions based on their colour and environment (see Fig.\ref{fig:colourmassdencolourandenv}). We observe that regions of red and blue galaxies are distinctly separated, which is consistent with our previous findings, as seen in the radial profiles. Furthermore, the distribution of points is influenced not only by colour but also by stellar mass density, since the division between regions of red and blue galaxies is traced by a diagonal line, and not just a vertical (stellar mass density) or horizontal (colour) line. Given the strong correlation between stellar mass density and radial distance \citep[see e.g.][and Sect.\ref{sec:radprof}]{Rosa2015,Rosa2016,Bluck2020}, we can consider stellar mass density as a proxy for the radial distance, at least to some extent. Consequently, these diagrams provide a different perspective on the colour profiles of galaxies.

No clear relation is found between galaxy environment and their position in these diagrams. The distribution of galaxies in groups is similar to that of galaxies in the field, with no noticeable clustering in specific regions. For a given mass-density bin, we find that the difference in the median $(u - r)\mathrm{int}$ colour of red galaxies between the field and groups is approximately $0.2$–$0.6$ times the median uncertainty on the $(u - r)\mathrm{int}$ colour of red galaxies in that bin. For blue galaxies, this difference is slightly larger, ranging from $0.4$ up to one times the typical uncertainty on the $(u - r)_\mathrm{int}$ colour in the same regions; however, it still remains within the associated uncertainties. In the highest density bins, this difference is larger (up to twice the typical error), but the low number of points can be playing an important role in this difference.  This result reflects our findings in the integrated analysis. In \cite{Julio2022}, we observed that the distribution of galaxy properties in the mass–colour diagram for the cluster mJPC2470-1771 was consistent with that of the general \mjp \ sample. The key difference was in the fraction of red and blue galaxies, which led to distinct clustering patterns. In contrast, our current results suggest that mass density and colour are more influential factors in the spatially resolved properties of galaxies than environment.

\begin{figure}
    \centering
    \includegraphics[width=0.45\textwidth]{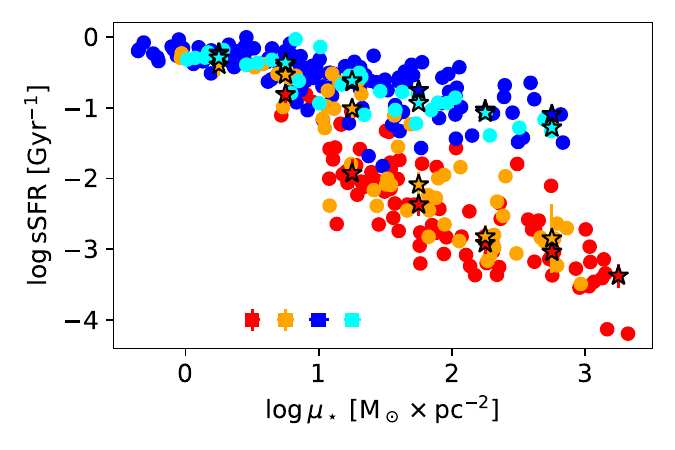}
    \caption[Local star formation main sequence]{Local star formation main sequence. Squares represent the typical error along each axis for each type of galaxy. Colour code is the same as in Fig~\ref{fig:colourmassdencolourandenv}. }
    \label{fig:localSFMS}
\end{figure}

We conclude this section of our analysis of stellar population properties by reproducing the local SFMS (see Fig.~\ref{fig:localSFMS}). Regions of red and blue galaxies are clearly separated. The difference of the sSFR for each mass density bin is usually $0.4$ to $0.7$ times the typical error for said bin, so we are not finding a significant effect of the environment on the blue galaxies in our sample in this point. Regarding red galaxies, we find that the difference for a given mass density bin is usually between the typical error and 4 times the typical error of that bin. This difference is statistically significant, but it is important to consider that, using the criteria by \cite{Peng2010}, this regions are already quenched (their sSFR is lower than $0.1$~Gyr$^-1$), so the actual value of the sSFR is of lesser relevance.  Our results are consistent with those of \cite{Rosa2016}, showing a strong relationship between the mass density of the region and its sSFR.

\subsection{Radial distribution of the stellar population properties} \label{sec:radprof}

\begin{figure*}
    \centering
    \includegraphics[width=\textwidth]{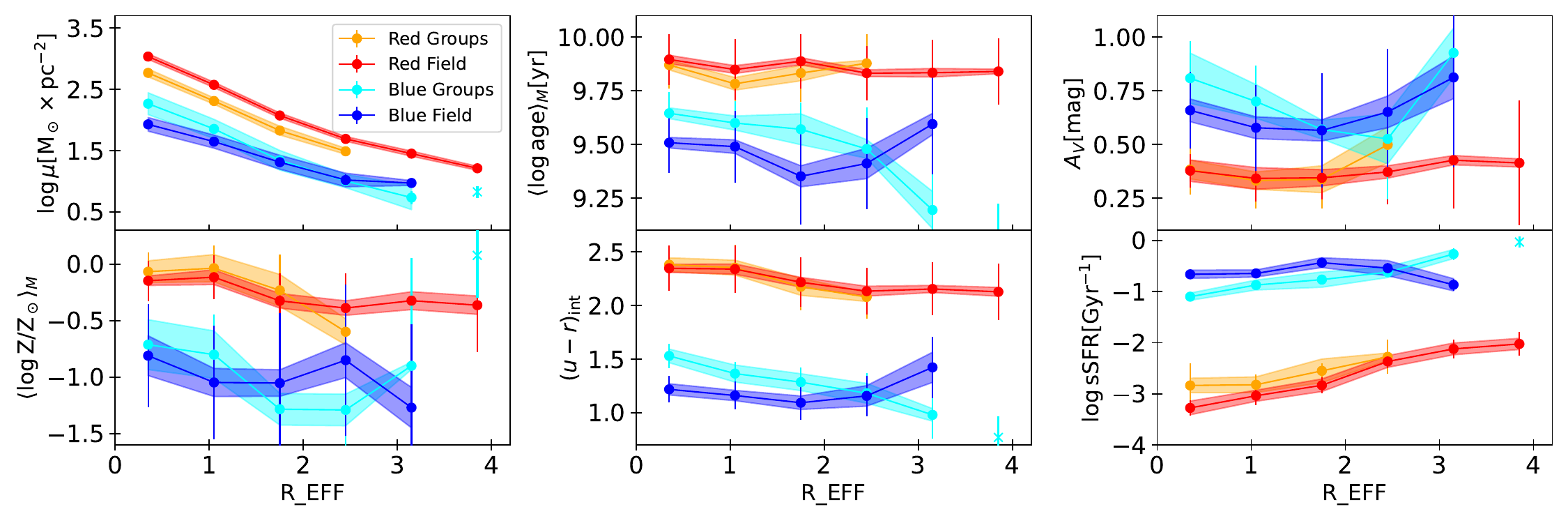}
    \caption[Radial profile and internal gradients of the of the different galaxy proerties by galaxy colour and environment.]{Radial profile and gradients of the stellar mass surface density by galaxy colour and environment. Colour code is the same as in Fig~\ref{fig:colourmassdencolourandenv} Dashed lines represent the median value in the radius bin. Colour shade represent the error of the median. Error bars represent the typical error in each bin. Single points represent bins where only one region was left after the S/N cleaning. }
    \label{fig:mass_Profile_gradient}
\end{figure*}

For the analysis of the radial distribution of the stellar population properties, we use a fixed-step segmentation (see  Fig.~\ref{fig:mass_Profile_gradient}). As in previous sections, we divide our sample into four groups based on environment and colour. Additionally, to facilitate comparison with existing literature, we assume that red galaxies are typically quiescent and early-type, while blue galaxies are generally star-forming and late-type \citep[see e.g.][]{Hogg2004,Kauffmann2004,Bassett2013}. Although these categories are not synonymous, they are strongly correlated, making them a useful proxy for comparison purposes, as different studies often classify galaxies using varying criteria. After close inspection, we found that 5 red galaxies in the sample could be considered spirals, and 1 blue galaxy could be considered an elliptical, but the PSF of the image makes the classification uncertain in some of these galaxies.

\subsubsection{Stellar mass surface density}

Stellar mass density (left top panel in  Fig.~\ref{fig:mass_Profile_gradient})  decreases with increasing galacto-centric radius for all galaxy spectral type and environments. Red galaxies in the field exhibit the highest densities, remaining above blue galaxies across all radii. At $0.7$~\texttt{R\_EFF}, the median stellar mass density drops to $\log \mu_\star \approx 3$, with values consistent with previous studies \citep[see e.g.][]{Rosa2015,Bluck2020,Abdurro2023,Ana2024} but slightly lower due to PSF homogenisation and lower resolution in our radial bins. Differences between studies are attributed to varying stellar initial mass functions (IMFs), as the Salpeter IMF used in \cite{Rosa2015} yields higher mass values than the \cite{Chabrier2003} IMF used here \citep[see also][]{Luis2024}.

Red galaxies in groups have densities similar to those in the field, but $\sim 0.1$~dex lower. This difference remains between 4 and 7 times larger than the typical error at each radii of the stellar mass surface density of red galaxies for all distances. These results align with \cite{Bluck2020}, who find that satellite galaxies in groups are less dense than central galaxies, though high dispersion in group galaxy profiles warrants further investigation. Blue galaxies in both the field and groups show similar profiles, consistent with lower mass galaxies in previous studies and with findings by \cite{Bluck2020} and \cite{Abdurro2023} for star-forming galaxies. The difference is between 3 to 8 times the typical error of the stellar surface mass density of blue galaxies in central parts, but is within the typical error at each radii for distances larger than $\sim 1$~\texttt{R\_EFF}.  

\subsubsection{Stellar ages}

Mass-weighted stellar ages (middle top panel in  Fig.~\ref{fig:mass_Profile_gradient}) exhibit similar profiles for red galaxies, whether in the field or in groups, remaining almost flat at $\left <\log \mathrm{age} \right>_M \approx 9.9$. Both types of red galaxies show a slight decrease in age up to 1~\texttt{R\_EFF}, followed by a slight increase, consistent with a rather flat profile. However, these trends fall within the uncertainty intervals, since the difference remains between $0.2$ and $0.5$ times the typical error at each radii of the age of the stellar ages of red galaxies, suggesting no significant environmental differences. Our findings align with previous studies that show a flat age profile for high-mass galaxies, consistent with \cite{Rosa2014, Rosa2015} and \cite{SanRoman2018}.

Blue galaxies, in contrast, show more noticeable differences in their age profiles. In the field, age decreases from $\left <\log \mathrm{age} \right>_M \approx 9.5$ at 1~\texttt{R\_EFF} to $\left <\log \mathrm{age} \right>_M \approx 9.4$ at 2~\texttt{R\_EFF}, then increases again at larger radii. Blue galaxies in the field tend to be slightly younger than those in groups, This difference is larger for distances lower than $\sim 2$~\texttt{R\_EFF}, where it is of the order 1 to $1.7$ times the typical error at each radii of the age of blue galaxies. At larger radii, this difference is around $0.3$ times the typical error, except for the last bin, where the lower number of point alters the final value. These results are consistent with previous studies, including \cite{Rosa2014, Rosa2015} and \cite{Bluck2020}, where blue galaxies show relatively flat age profiles, with younger populations at larger radii. We also observe that blue galaxies in groups are generally slightly more massive, but their age profiles remain similar across environments.

\subsubsection{Extinction}

Extinction (right top panel in  Fig.~\ref{fig:mass_Profile_gradient}) shows a generally flat profile for red galaxies, both in groups and in the field, at around $A_V \approx 0.4$. In the field, red galaxies show a slight increase in $A_V$ from $\sim 2$~\texttt{R\_EFF}, but due to limited data at larger radii, this trend cannot be firmly confirmed. Overall, no significant environmental effect on extinction is observed, likely due to measurement uncertainties. The difference remains between $0.04$ and $0.8$ times the typical error at each radii of the extinction of red galaxies. Our results are consistent with \cite{Rosa2015}, who observed a steep decrease in extinction up to $\sim 0.5$~\texttt{R\_EFF}, followed by a flatter gradient. These findings also align with \cite{SanRoman2018}, who reported small positive gradients, with $A_V \approx [0.2, 0.5]$.

For blue galaxies, extinction decreases with radius in groups, from $A_V \approx 0.9$ in the central regions to $A_V \approx 0.5$ at $\sim 2.5$~\texttt{R\_EFF}. This decrease correlates with the steep drop in mass-weighted age found for blue galaxies in groups, although the lack of data at larger radii suggests a possible degeneracy between age and extinction. In the field, extinction decreases from $A_V \approx 0.75$ at the centre to $A_V \approx 0.5$ at $\sim 2$~\texttt{R\_EFF}, but without a steep change in age. While blue galaxies in groups tend to show slightly higher extinction, this difference is minimal between $0.02$ and $0.9$ times the typical error at each radii of the extinction of blue galaxies and may be related to mass distribution, as suggested by \cite{Rosa2015}. Our results show a similar pattern, with extinction decreasing in the inner regions of blue galaxies, generally ranging from $A_V \approx [0.4, 0.6]$, consistent with \cite{Rosa2015}, although we observe slightly higher extinction values.

\subsubsection{Metallicity}

Stellar mass-weighted metallicity profiles (left bottom panel in  Fig.~\ref{fig:mass_Profile_gradient}) decrease with radius for all galaxy types. For red galaxies in the field, metallicity drops from $\left < \log \mathrm{Z/Z_\odot} \right > \approx -0.25$ in the central regions to $\left < \log \mathrm{Z/Z_\odot} \right > \approx -0.5$ at $\sim 2$~\texttt{R\_EFF}, then remains constant in outer regions. Red galaxies in groups show a flatter metallicity profile up to $\sim 2$~\texttt{R\_EFF}, though the uncertainty intervals make it difficult to confirm if environmental effects influence these profiles. These differences remain between $0.3$ and $0.7$ times the typical error at each radii of the metallicity of red galaxies. compared with \cite{Rosa2014, Rosa2015}, which report a modest decrease in metallicity, our results show more metal-poor galaxies, consistent with the findings of \cite{SanRoman2018}. The discrepancies in metallicity are within expected uncertainty intervals and could arise from differences in methodology.

For blue galaxies, the metallicity profile is similar for both group and field galaxies, decreasing from $\left < \log \mathrm{Z/Z_\odot} \right > \approx -0.5$ at the centre to $\left < \log \mathrm{Z/Z_\odot} \right > \approx -1.25$ at $\sim 3$~\texttt{R\_EFF}, with the drop potentially influenced by a degeneracy between stellar age, extinction, and metallicity. In the field, the profile shows a break around $\sim 2.5$~\texttt{R\_EFF}, where the metallicity decrease slows. The difference between the metallicities of blue galaxies in the field and in groups remains between $0.2$ and $0.7$ times their typical error at each radii. If we compare our results with those by \cite{Rosa2014,Rosa2015}, focusing in their galaxies within the mass range $\log M_\star = [9.1,10.6]$~$[M_\odot]$, we find that the metallicity decreases with distance too, although at a slower rate. In fact, their galaxies with masses $\log M_\star < 10.1$~$[M_\odot]$, the profile looks rather flat, which results in a good agreement with our results.

\subsubsection{Colour}

The $(u-r)_\mathrm{int}$ colour profiles (middle bottom panel in  Fig.~\ref{fig:mass_Profile_gradient}) clearly differentiate red and blue galaxies. For red galaxies, the profiles are similar in both the field and groups, showing no significant environmental difference. In the central regions, red galaxies have $(u-r)_\mathrm{int} \approx 2.4$, decreasing to $(u-r)_\mathrm{int} \approx 2.1$ at $\sim 1.5$~\texttt{R\_EFF}, after which they continue to decrease. In the field, the decrease begins at $\sim 1.2$~\texttt{R\_EFF}, followed by a plateau. However, due to uncertainty intervals and differences in the number of galaxies in each environment, the environmental effect on colour is unclear. Nonetheless, the difference of their values remain between $0.06$ and $0.26$ times the typical error at each radii of the $(u-r)_\mathrm{int}$ colour for red galaxies. 

For blue galaxies, the colour profiles show more noticeable differences between the field and groups. In groups, $(u-r)_\mathrm{int}$ decreases from $\approx 1.55$~mag, becoming bluer more rapidly. In contrast, blue galaxies in the field exhibit a much flatter profile, with $(u-r)_\mathrm{int} \approx 1.25$~mag at the centre, decreasing slightly to $\sim 1.2$mag at $\sim 2$~\texttt{R\_EFF}, then increasing to $\approx 1.5$mag at $\sim 3$~\texttt{R\_EFF}. This reddening at larger radii may be attributed to the fact that these outer regions can only be measured with sufficiently high signal-to-noise ratios in more massive galaxies, which tend to be redder and more heavily obscured. The difference in colour between blue galaxies in the field and in groups, which decreases from $2.5$ to $0.25$ times the typical error of $(u-r)_\mathrm{int}$ at each radii, is not large enough to confidently link it to environmental effects, given the mass distribution of blue galaxies in the field, which skews toward more massive galaxies.

\subsubsection{Specific star formation rate}

We observe a clearly bimodal distribution in the radial profiles of the sSFR (right bottom panel in  Fig.~\ref{fig:mass_Profile_gradient}), dividing galaxies into red and blue populations. For red galaxies, the sSFR slightly increases with radial distance, both in the field and in groups. The profiles mostly overlap within the uncertainty, with differences ranging from 1 to 2 times the typical error at each radii of the sSFR of red galaxies. Values increasing from $\log \mathrm{sSFR} \approx -3$ in the central regions to $\log \mathrm{sSFR} \approx -2$ at 4~\texttt{R\_EFF} for field galaxies and $\log \mathrm{sSFR} \approx -2.5$ at $2.5$~\texttt{R\_EFF} for group galaxies. These profiles are consistent with those found by \cite{Rosa2016} for elliptical and S0  galaxies, and \cite{Abdurro2023} for quiescent galaxies. Red galaxies in both environments remain below the $\log \mathrm{sSFR} = -1$ threshold adopted by \cite{Peng2010} for separating star-forming galaxies from quiescent systems, with negligible environmental differences. 

Blue galaxies in the field exhibit flatter sSFR profiles at $\log \mathrm{sSFR} \approx -0.85$, while those in groups show an increase from $\log \mathrm{sSFR} \approx -1$ in the centre to $\log \mathrm{sSFR} \approx 0$ at 4~\texttt{R\_EFF}. This increase may be flatter when considering other properties. Both blue galaxy populations exceed the $\log \mathrm{sSFR} = -1$ threshold from \cite{Peng2010}, indicating active star formation. The sSFR of blue galaxies in the field appears slightly higher than in groups, suggesting possible quenching in groups, with differences decreasing from around 9 times to $0.7$ times the typical error at each bin of the sSFR of blue galaxies. These trends are consistent with previous studies, such as \cite{Rosa2016} for Sc and Sd galaxies, and \cite{Abdurro2023} for star-forming galaxies and those in the green valley. Our results align closely with these studies, while \cite{Ana2024} report quenched sSFR in the inner regions of high-mass spirals, which matches the flattening observed in our data.

\subsection{Emission lines}\label{sec:res:EL}

In this section, we study the main predictions of the ANN for the regions we have obtained with our methodology. We use the homogeneous rings segmentation for the same reasons as in the previous section, i.e., in order to prevent higher resolution galaxies from dominating the radial profiles and mass density-colour diagrams. We also analyse the BPT \citep{BPT} and WHAN \citep{WHAN_1,WHAN_2} diagrams and separate star--forming regions and regions hosting and AGN.

\subsubsection{Line emission of the regions}
\begin{figure*}
    \centering
    \includegraphics[width=\textwidth]{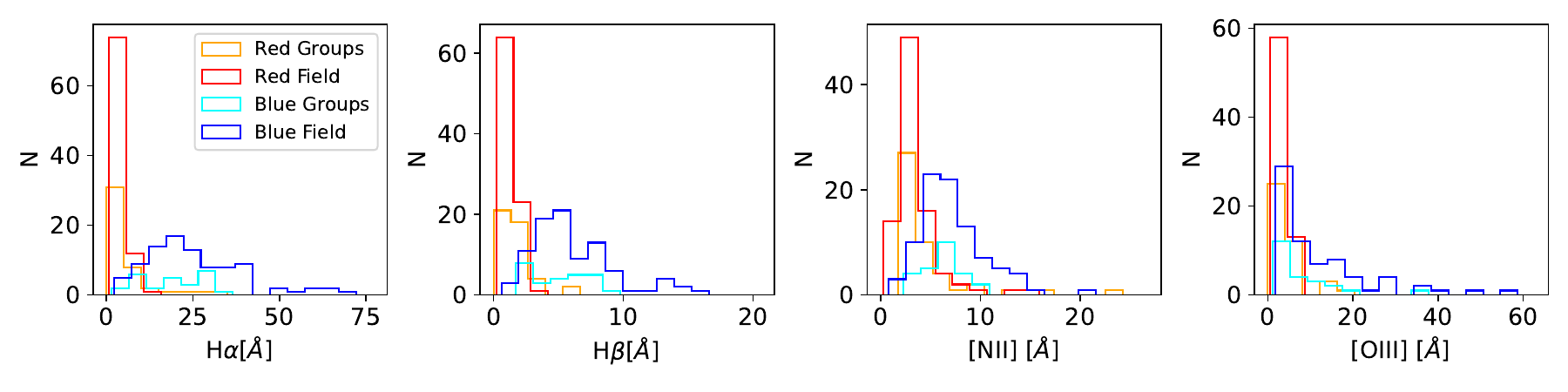}
    \caption[Histograms of the EW of the line emission of the regions of our sample of galaxies, by colour and environment]{Histograms of the EW of the line emission of the regions of our sample of galaxies, by colour and environment. From left to right: $\mathrm{EW(H\alpha)}$, $\mathrm{EW(H\beta)}$, $\mathrm{EW([NII])}$, and $\mathrm{EW([OIII])}$. Colour code is the same as in previous figures.}
    \label{fig:EWhist}
\end{figure*}

\begin{figure}
    \centering
    \includegraphics[width=0.4\textwidth]{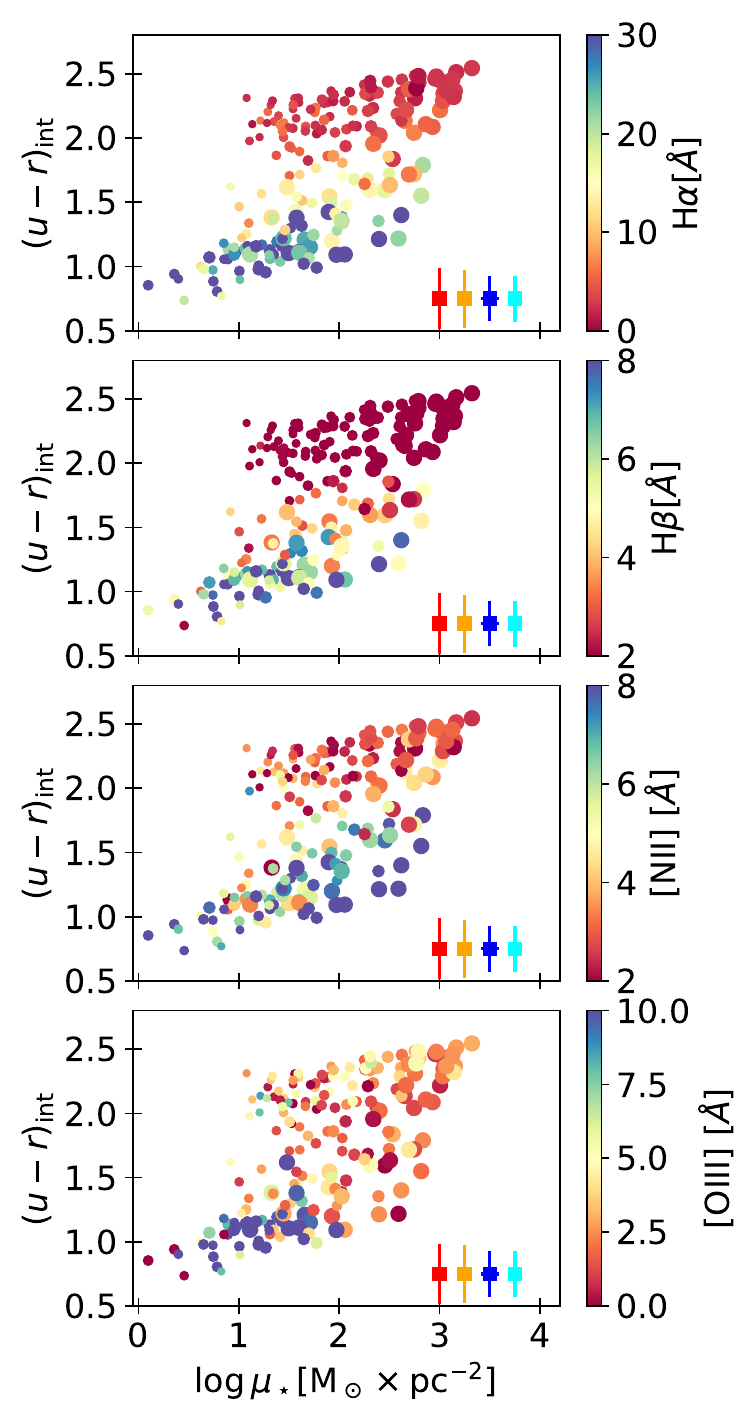}
    \caption[Colour--mass density diagram coloured by the main stellar population properties.]{Colour--mass density diagram coloured by the main stellar population properties.  From  top to bottom: $\mathrm{EW(H\alpha)}$, $\mathrm{EW(H\beta)}$, $\mathrm{EW([NII])}$, $\mathrm{EW([OIII])}$. Squares represent the typical error along each axis colour for each type of galaxy.}
    \label{fig:colourmassdenEW}
\end{figure}

We begin our analysis of the emission lines by examining the distributions of equivalent widths (EWs) and line emission ratios across galaxy regions in our sample, classified by colour and environment (see Fig.~\ref{fig:EWhist}). We note that the median error for these EWs are $4$,$0.8$,$1.5$, and $2.76$~\AA \ for $\mathrm{H\alpha}$, $\mathrm{H\beta}$, $\mathrm{[NII]}$, and $\mathrm{[OIII]}$. Given the limit of detection expressed by Eq.~6 from \cite{Gines2021} \citep[see also][]{Pascual2007}, most regions with EW lower than $7.7$, 9, $7.3$, and $8.3$~\AA  are compatible with no emission.

Red galaxies, both in the field and in groups, exhibit very low H$\alpha$ emission, with $\mathrm{EW(H\alpha)} < 10$~\AA. In contrast, blue galaxies show significantly stronger and more varied H$\alpha$ emission. For blue galaxies in groups, $\mathrm{EW(H\alpha)}$ spans from a few \AA\ up to $\sim 30$~\AA, whereas in the field it extends to nearly $\sim 75$~\AA, peaking at around 20 \AA. The environment appears to have little effect on red galaxies, and the differences seen in blue galaxies may stem from variations in sample size. These trends are consistent with known correlations between stellar mass and $\mathrm{EW(H\alpha)}$ \citep[e.g.][]{Fumagalli2012, Sobral2014, Khostovan2021}. As $\mathrm{EW(H\alpha)}$ traces star formation rate (SFR) \citep{Kennicutt1998,Kennicutt2012,Marmolqueralto2016,Khostovan2021}, the higher values observed in blue galaxies align with their star-forming nature \citep[e.g.][]{Peng2010,Bluck2014} and our own findings of elevated SFR intensities in these regions.

Similar patterns are observed for H$\beta$. Red galaxies across environments display weak emission ($\mathrm{EW(H\beta)} \approx 0$–3~\AA), whereas blue galaxies exhibit stronger emission. In groups, blue galaxies show $\mathrm{EW(H\beta)}$ up to $\sim 10$~\AA, and in the field up to $\sim 17$~\AA, with a peak at 5~\AA.

The distribution of [NII] emission shows subtle differences. Red galaxies still have relatively low $\mathrm{EW([NII])}$ values, though many regions have values above 0~\AA, reaching up to $\sim 15$~\AA. Blue galaxies generally show higher [NII] emission: group galaxies range from $\sim 1$ to 10 \AA, while field galaxies extend up to 15~\AA, with both distributions peaking at around 5~\AA.

Finally, [OIII] emission also follows this pattern: red galaxies show weak emission (0–10~\AA), while blue galaxies exhibit stronger [OIII] emission, particularly in the field (up to 60~\AA) and slightly lower in groups (up to 20~\AA).

The mass density–colour diagram (see Fig.~\ref{fig:colourmassdenEW}) provides a clear interpretation of the data, as both the EWs and line ratios exhibit well-defined distributions across the diagram. There is a strong correlation with galaxy colour, particularly in the case of the EWs of H$\alpha$ and H$\beta$. While this may be partly attributed to the fact that the ANNs are trained using colour information, there is also well-established evidence linking emission line strength to galaxy colour classification  \citep[i.e. blue vs. red galaxies, see e.g.][and references therein]{Gines2022}. Nonetheless, the influence of stellar mass surface density on emission properties should not be underestimated.

We observe that the equivalent width of emission lines increases in blue regions with low stellar mass surface density. These regions also show the greatest variability in EW values. Based on earlier arguments using mass density as a proxy for galactocentric distance, these areas correspond predominantly to the outer regions of blue galaxies. This variability is expected, as galaxies differ in their emission characteristics and include populations of extreme emission line galaxies \citep[see e.g.][and references therein]{Iglesias2022, Gines2022}. Conversely, as regions become redder—particularly those with higher stellar mass density—the EWs decrease markedly. These are typically the inner regions of red galaxies, which are generally quiescent and exhibit very low emission. The equivalent width of [OIII], however, is only significant in low-density, blue regions, and diminishes rapidly elsewhere. It is important to note that our estimation of [OIII] emission is subject to greater uncertainty \citep{Gines2021}.

\subsubsection{WHAN and BPT diagrams}

\begin{figure*}
    \centering
    \includegraphics[width=\textwidth]{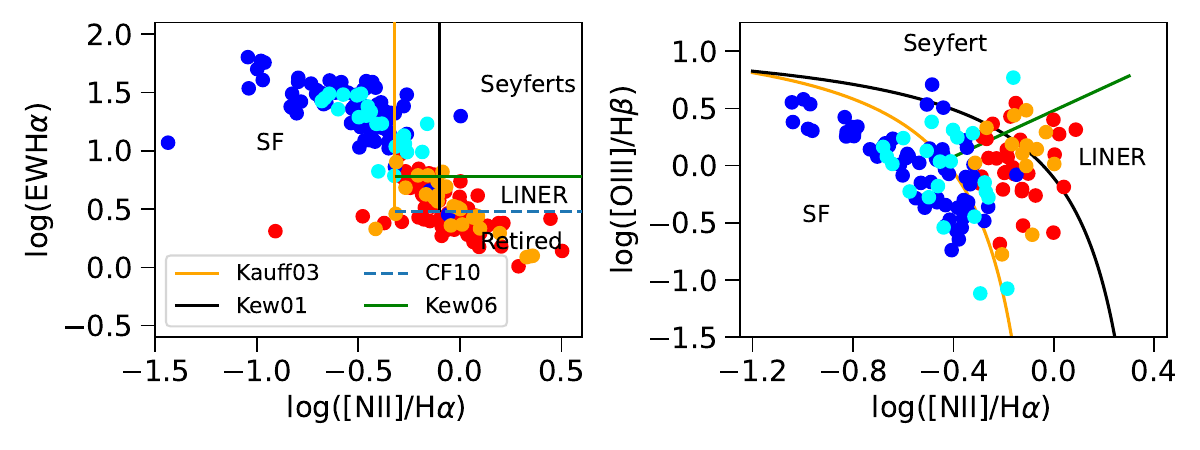}
    \caption[WHAN and BPT diagrams of the regions of the spatially resolved galaxies in \mjp]{WHAN (left panel) and BPT (right panel) diagrams of the regions of the spatially resolved galaxies in \mjp. Red points represent red galaxies in the field. Orange points represent red galaxies in groups. Blue points represent blue galaxies in the field. Cyan points represent blue galaxies in groups. }
    \label{fig:WHANspatially}
\end{figure*}

We now analyse the BPT \citep{BPT} and WHAN \citep{WHAN_1, WHAN_2} diagrams to classify regions as star-forming or AGN-dominated, using our segmentation based on homogeneous ring analysis (see Fig.~\ref{fig:WHANspatially}). We note that in the BPT, we only show regions above the retired line in the WHAN diagram, in order to avoid biases and missclassifications produced by unreliable estimations in low or no emission line regions. Regions belonging to blue galaxies are generally located within the star-forming area of both diagrams, although some fall into the “composite” zone, lying between the demarcation lines proposed by \citet{Kauffmann2003} and \citet{Kewley2001}. Only a small number of these regions appear in the LINER part of the diagrams, and we identify just one potential Seyfert region in the WHAN diagram and three in the BPT diagram.

In contrast, regions associated with red galaxies are predominantly found in the LINER and “retired” areas of the WHAN diagram. In the BPT diagram, these regions tend to lie in the “composite” zone, with a greater number classified as Seyferts. It is important to note the lower accuracy in our estimation of the $\mathrm{[OIII]/H\beta}$ ratio, which affects the classification in the BPT diagram. Furthermore, we are unable to distinguish between galaxies in groups and those in the field based on these diagnostics.

            \begin{table}[]
        \centering
        \begin{tabular}{|c|c|c|c|c|}
        \hline
           Class & RG & RF & BG & BF \\
           \hline
             SF & $0$~\%  & 0~\% & $66.67$~\% & $90.47$~\% \\
             Composite & $44.44$~\% & $33.33$~\% &  $33.33$~\% & $4.76$~\% \\
            Seyfert & $0$~\% & $0$~\% & $0$~\% & $0$~\% \\
            LINER & $11.11$~\% & $6.67$~\% & $0$~\% & $0$~\% \\
            Retired & $44.4$~\% & $60$~\% & $0$~\% & $0$~\% \\
            \hline
        \end{tabular}
            \caption{Classification of the central regions of galaxies in the sample, attending to the WHAN diagram. Galaxies are divided by their spectral type and environment: red galaxies in groups (RG), red galaxies in the field (RF), blue galaxies in groups (BG) and blue galaxies in the field (BF).}
        \label{tab:AGN_frac}
        \end{table}

To further investigate the role of the environment in triggering nuclear activity, we analyse the classifications obtained from the WHAN diagram for the innermost region of each galaxy (see Table~\ref{tab:AGN_frac}). We focus on the WHAN diagram since it offers a better classification for low emission galaxies given our filter system \citep[see][]{Gines2021,Gines2022}. We use the maximum-resolution segmentation to delineate regions that enclose the innermost part of the galaxy. As before, these are classified according to their spectral type and environment.

We find that the percentage of AGN-host regions does not vary significantly between red galaxies in groups and red galaxies in the field. Conversely, we observe a higher percentage of nuclei in the composite regions of the diagrams for blue galaxies in groups compared with those in the field. Results from literature on this subject are mixed. For instance, \cite{Peluso2022} report a strong correlation between AGN activity and ram-pressure stripping, which is more intense in denser environments, based on their optical sample. However, \cite{Tiwari2025} find no such correlation using an X-ray-based sample. Other studies, such as those by \cite{Dressler1985}, \cite{Kauffmann2004}, and \cite{Lopes2017}, observe a higher presence of AGNs in lower-density environments, while works like those by \cite{Amiri2019} and \cite{MunozRodriguez2024} do not find any significant relation with environment. In contrast, \cite{deVos2024} report a larger fraction of AGNs in the outskirts of clusters and in the very central regions, with a lower fraction in intermediate regions. Our previous work \citep{Julio2022} shows a higher fraction of AGN hosts in the core of the most massive cluster in \mjp. Our results suggest a scenario in which the environment does not play a significant role for red galaxies, but does for blue galaxies. However, the scope of our conclusions is limited by the small sample size. A small statistical variation can lead to a considerable fluctuation in the reported AGN-host fraction. Therefore, our results should be regarded as proof of concept and not be generalised, as a larger sample would be necessary to either confirm or challenge this hypothesis.

\subsection{SFH}\label{sec:SRSFH}

\begin{figure}
    \centering
    \includegraphics[width=0.45\textwidth]{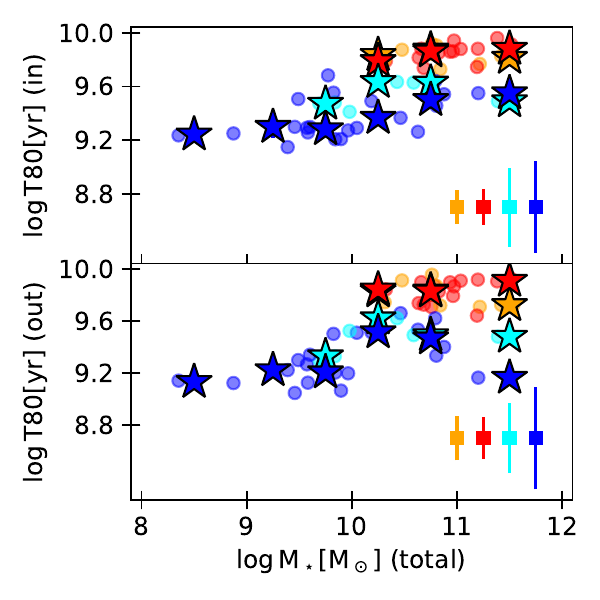}
    \caption[$\mathrm{T80}$ vs galaxy total mass]{$\mathrm{T80}$ vs galaxy total mass. The upper panel shows the values for the inner region, while the bottom panel presents the values for the outer region. Colour code is the same as in previous figures. Stars represent the median value for each type of galaxy in each stellar mass bin. Squares represent the typical error along each axis colour for each type of galaxy.}
    \label{fig:T80}
\end{figure}

We finish our section of results by studying the SFH of the galaxies. For this section, we use the inside-out segmentation, in order to compare the SFH of the inner and outer regions. We study the parameter $\mathrm{T80}$, which is the loockback time at which the galaxy has formed 80~\% of its stellar mass (taking into account the mass loss due to stars reaching the end of their lifetime). We show this parameters as a function of the total stellar mass, colour coded by the colour of the galaxy and its environment, in Fig.~\ref{fig:T80}.

We find a strong correlation between $\mathrm{T80}$ and stellar mass. Red, more massive galaxies exhibit higher $\mathrm{T80}$ values in both inner and outer regions, indicating that these galaxies formed the bulk of their stars at earlier cosmological times. The similarity of $\mathrm{T80}$ between inner and outer regions suggests that star formation occurred over comparable timescales throughout the galaxy, within the precision limits of our stellar population models.

Conversely, blue galaxies display lower $\mathrm{T80}$ values, suggesting more recent star formation. These galaxies also show greater dispersion in the $\mathrm{T80}$ values of their inner regions, which tend to be slightly higher than those in the outer regions. These differences are, however, really small, around $0.04$~dex for red galaxies and $0.07$~dex for blue galaxies. These values are around $0.1$ times the typical error of T80 for red galaxies, and $0.2$ times the typical error of T80 for blue galaxies. This result implies that blue galaxies may have formed their inner parts, possibly bulges, earlier than their outer discs, consistent with an inside-out growth scenario, as reported in several previous studies \citep[see e.g.][]{MunozMateos2007, Perez2013, 
IbarraMedel2016, Zheng2017, Ruben2017}, although the error budget is also compatible with a similar SFH for both parts.

The environment does not play a significant role in the differences in $\mathrm{T80}$ between red galaxies, neither in the inner nor in the outer regions. For blue galaxies, we find some discrepancies: the inner regions of intermediate-mass blue galaxies ($\sim10^{10}\ \mathrm{M}_\odot$) in groups appear to have formed earlier than those of similar-mass galaxies in the field.

\section{Discussion} \label{sec:discuss}

\begin{figure*}
    \centering
    \includegraphics[width=\textwidth]{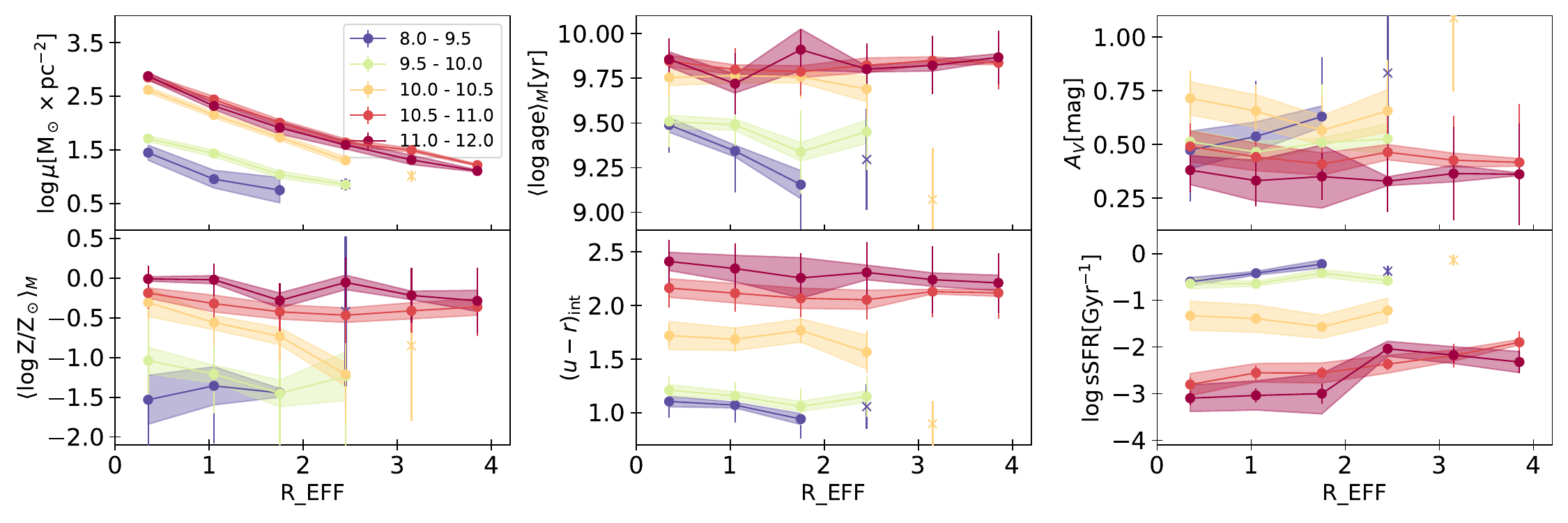}
    \caption[Radial profiles of the stellar population properties, divided by mass bins.]{Radial profiles of the stellar population properties, divided by mass bins. From left to right, top to bottom: stellar mass surface density, mass-weighted age, extinction, stellar metallicity, $(u-r)_\mathrm{int}$ colour, and sSFR. Different colours represent mass bins. Dashed lines represent the median value in each mass bin, shades represent the error of the median, error bars represent the typical error in each bin, and single points represent radius bins with only one region. }
    \label{fig:radialprofmass}
\end{figure*}

Throughout this paper, we have found radial profiles of the stellar population properties that are in agreement with the literature, particularly (but not exclusively) the works by \cite{Rosa2014,Rosa2015,Rosa2016,SanRoman2018,Bluck2020,Parikh2021} and \cite{Abdurro2023}, and, from a certain perspective, our results mirror the well-established integrated properties of galaxies. Red galaxies are generally more massive, older, more metal-rich, and exhibit lower SFR and sSFR than blue galaxies. Our results show that this behaviour also holds on local scales: regions belonging to red galaxies are typically redder, older, more metal-rich, have higher stellar mass surface densities, and display lower sSFR than regions in blue galaxies at the same galactocentric distance. The profiles show clear differences between red and blue galaxies, but no significant difference between galaxies in the field and in groups. We note here that we are limited by several factors. The first one and probably the most important of these is that our sample of galaxies is quite small. The results themselves for each galaxy are trustworthy because of our selection and because of the proofs performed in \cite{Julio2025}, but their statistical significance will improve once more data become available. Another factor is that our selection of galaxies is limited to galaxy groups, which usually have lower masses than clusters. This is important because the efficiency of several environment related process increases with the mass of the group or cluster \citep{Alonso2012, Raj2019}. In particular, the range of the total stellar mass of groups, which is related to total mass is $\sim 10^{10.25}$--$10^{11.5}$~$M_\odot$ for groups containing our blue galaxies, and $\sim 10^{10.75}$--$10^{12}$~$M_\odot$ for groups containing our red galaxies. This mass range is significantly lower than the total stellar mass of the galaxy cluster studied in \cite{Julio2022}, which is around $10^{12.5}$~$M_\odot$.  Related to this point, in \cite{Rosa2022} we found differences between the integrated properties of galaxies in groups and in the field, but many of these differences were due to the fraction of red and blue galaxies, and the properties of galaxies with the same mass and colour also showed similar properties, regardless of their environment, similarly to the results found here. Lastly, the resolution of the binning of the galaxy affects the profile: a lower spatial resolution results in a flatter profile. Therefore, we may be able to produce profiles with better resolution showing more significant differences in the future, when data of galaxies with larger apparent sizes are available.

In general, we find that, due to their mass, red and blue galaxies are largely well separated in the various diagrams analysed in this paper. However, similarly to the case of the radial profiles, we observe no significant differences between galaxies in groups and field. Regarding the results related to the gradients of the stellar population properties, the two galaxies in our sample with masses below $10^9$~$M_\odot$ are distinct from the rest and, when considered as outliers, we can identify some trends in the gradients in relation to galaxy stellar mass.

Concerning the role of environment on galaxy evolution, it is common to compare it to the role of the stellar mass, as reflected by the division of the quenching process into mass quenching and environmental quenching \citep[see e.g.][]{Peng2010,Ilbert2013}. In fact, the radial profiles of stellar population properties studied by \cite{Rosa2015} show a dependence not only on the morphological type of the galaxy, but also on its total stellar mass, as well as the galaxies studied by \cite{Ana2024}. Furthermore, the work by \cite{Zibetti2022} points to the stellar mass both as local and global driver of the evolution of galaxies. For this reason we decide to include the radial profiles of the stellar population properties previously studied, but divided into mass bins (see Fig.~\ref{fig:radialprofmass}). 

We find that, for most properties, their values are well differentiated by mass at all distances. In particular, the stellar mass surface density is always higher for massive galaxies than for low mass galaxies, and massive galaxies are older, more metal-rich, and redder at all distances compared to low-mass galaxies. Additionally, the profiles of the sSFR also show clear offsets with mass, where low mass galaxies have significantly higher sSFR than massive galaxies. The most notable exception is extinction, which shows similar profiles across all masses. On the other hand, the radial profiles of the intensity of the SFR are almost bimodal, with galaxies in the range $M_\star = [10^{8},10^{10.5}]$~$M_\odot$ showing comparable profiles within the uncertainty intervals, but still clearly differentiated from galaxies with masses larger than $10^{10.5}$~$M_\odot$. These results show that the mass indeed plays a significant role in the determination of the local properties of galaxies, as found in the aforementioned works.

However, we also point out that the mass of our groups may be too low for the environment to exert any significant effect in our spatially resolved galaxies. Indeed, the mass of the groups in \mjp \ is very low in comparison to massive clusters, as shown by \cite{Rosa2022}. In that work, we showed how the quenched fraction excess of galaxies was significantly higher for the cluster mJPC2470-1771 than for low mass groups. Therefore, we may see more significant effects in the properties of the spatially resolved galaxies in groups and clusters with higher mass in the future data releases of \mjp.

\section{Summary and conclusions}
We have applied our tool, \PyDJ, to the spatially resolved galaxies from the \mjp \ survey in order to study the effect of the environment in the spatially-resolved  properties of 51 galaxies in \mjp. 
We have checked the observational and the integrated properties  of the sample, checking that red galaxies in the field and in groups are comparable between them, as well as blue galaxies in the field and in groups, but we must take into account that the sample of blue galaxies in the groups is biased towards higher masses. 

In the spatially resolved analysis, we have mainly used elliptical rings of semi-major axis the size of the FWHM of the worst PSF for each galaxy, and elliptical rings of $0.7$~\texttt{R\_EFF}.  
We have studied the stellar populations and emission lines properties of these regions, using a mass density--colour diagram and their radial profiles and gradients. Lastly, we compare the SFH inside-out of galaxies, using elliptical rings defined from the centre up to $0.7$~\texttt{R\_EFF} for inner regions, and from $0.7$~\texttt{R\_EFF} to $2.5$~\texttt{R\_EFF}. Our main conclusions are:

\begin{itemize}
    \item Our tool, \PyDJ, provides solid magnitude measurements that offer reliable galaxy properties.
    \item The  properties of the regions are distributed clearly in the mass density-colour diagrams, similarly to the integrated mass-colour diagram. We find that redder, denser regions are usually older, more metal rich, and show lower values of the sSFR (they are more quiescent) than bluer, and less dense regions. The highest value of the extinction $A_V$ are found in blue, dense regions, as well as some of the most metal rich regions. The regions of red and blue galaxies remain clearly separated in these diagrams.
    \item The radial profiles of the properties of the galaxies that we obtain are compatible with an extensive literature 
    The profiles of the red and blue galaxies are clearly different, but we do not find any remarkable effect of the environment.    
    \item 
    The equivalent widths (EWs) of emission lines vary across the mass density–colour diagram, with the highest EWs in the bluest, least dense regions. WHAN and BPT diagrams show that blue galaxy regions are mostly star-forming, whereas red galaxy regions are classified as LINER or retired. 
    \item The comparison of the SFH of the inner and outer regions suggests and inside-outside formation scenario.
    \item We find that in general, the properties of the regions of red and blue galaxies are well distinguished, but that there is no significant effect of the environment in the properties of these regions. 
\end{itemize}

\begin{acknowledgements}
J.E.R.M., L.A.D.G., R.M.G.D., G.M.S., R.G.B., and I.M. acknowledge financial support from the Severo Ochoa grant CEX2021-001131-S funded by MICIU/AEI/ 10.13039/501100011033.
J.E.R.M., L.A.D.G., R.M.G.D., G.M.S., and R.G.B., are also grateful for financial support from the project PID2022-141755NB-I00, and proyect ref. AST22\_00001\_Subp 12 and 11 with fundings from the European Union - NextGenerationEU»; the Ministerio de Ciencia, Innovación y Universidades ; the Plan de Recuperación, Transformación y Resiliencia ; the Consejería de Universidad, Investigación e Innovación from the Junta de Andalucía and the  Consejo Superior de Investigaciones Científicas.
I.M. acknowledges financial support from the project PID2022-140871NB-C21
Based on observations made with the JST/T250 telescope and PathFinder camera for the miniJPAS project at the Observatorio Astrofísico de Javalambre (OAJ), in Teruel, owned, managed, and operated by the Centro de Estudios de Física del Cosmos de Aragón (CEFCA). We acknowledge the OAJ Data Processing and Archiving Unit (UPAD) for reducing and calibrating the OAJ data used in this work. Funding for OAJ, UPAD, and CEFCA has been provided by the Governments of Spain and Aragón through the Fondo de Inversiones de Teruel; the Aragón Government through the Research Groups E96, E103, E16\_17R, and E16\_20R; the Spanish Ministry of Science, Innovation and Universities (MCIU/AEI/FEDER, UE) with grant PGC2018-097585-B-C21; the Spanish Ministry of Economy and Competitiveness (MINECO/FEDER, UE) under AYA2015-66211-C2-1-P, AYA2015-66211-C2-2, AYA2012-30789, and ICTS-2009-14; and European FEDER funding (FCDD10-4E-867, FCDD13-4E-2685).
This paper has gone through internal review by the J-PAS collaboration. We thank Gustavo Bruzual for his work as an internal referee, as well as Iris Breda and Álvaro Álvarez Candal for their comments to improve the paper. 
\end{acknowledgements}

   \bibliographystyle{aa} 
   \bibliography{main} 

\appendix

\section{Observational and integrated properties of the sample} \label{app:properties}

\begin{figure*}
    \centering
    \includegraphics[width=\textwidth]{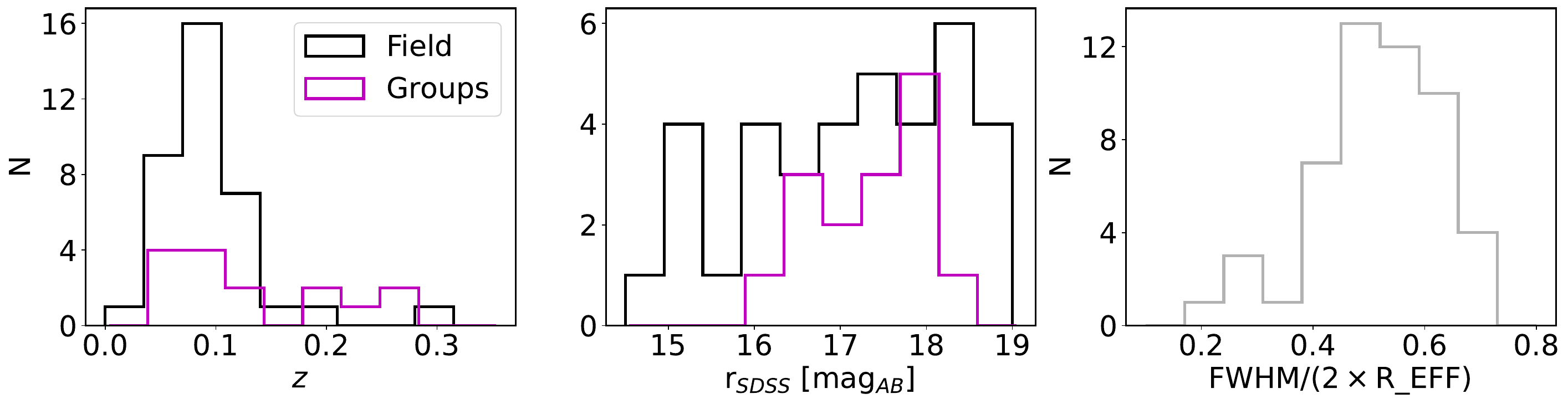}
    \caption[Integrated observational properties of the spatially resolved galaxies in \mjp]{Integrated observational properties of the spatially resolved galaxies in this work. First panel shows the histogram of the redshift of the galaxies. Middle panel shows the distribution of the magnitude in the \rb{} of the \magauto \ photometry. Black histograms represent galaxies in the field, and magenta histograms represent galaxies in groups. Last panel shows the comparison of the FWHM of the worst PSF for each galaxy and the double of its effective radius for the complete sample.}
    \label{fig:intobsprop}
\end{figure*}

\begin{figure*}
    \centering
    \includegraphics[width=\textwidth]{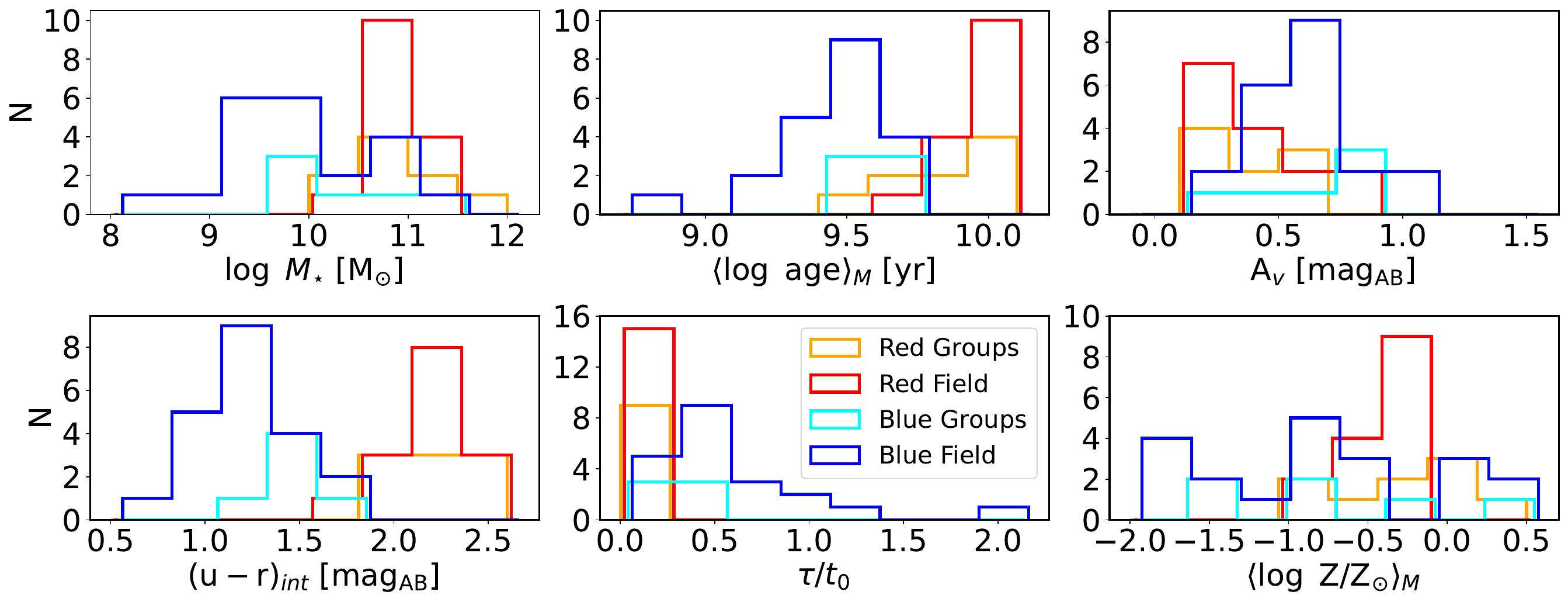}
    \caption[Comparison of the integrated stellar population properties properties of the spatially resolved galaxies by colour and environment]{Comparison of the integrated stellar population properties properties of the spatially resolved galaxies by colour and environment. From left to right, top to bottom: stellar masss, mass-weighted age, extinction $A_V$, $(u-r)_\mathrm{int}$, $\tau/\mathrm{t}_0$, and stellar metallicity. Colour code is the same as in previous figures}
    \label{fig:intSPP}
\end{figure*}

In this appendix we summarise the main observational (Fig.~\ref{fig:intobsprop}) and integrated (Fig.~\ref{fig:intSPP}) properties of the sample of galaxies. We compare these properties to avoid biases and wrong conclusions in our analysis when analysing the role of environment on galaxy evolution.

The most important aspect of the comparison of the observational properties (Fig.~\ref{fig:intobsprop}) is that galaxies in groups and in the field cover a similar redshift range, which ensures a fair comparison between properties of both types of galaxies. We also find that there are brighter and also dimmer galaxies in the field than in groups, but this is most likely due to the different size of the samples. Lastly, the last panel of Fig.~\ref{fig:intobsprop} shows the comparison of the \texttt{R\_EFF} of each galaxy to the FWHM of its worst PSF, justifying our choice of steps of $0.7$~\texttt{R\_EFF} during our analysis.

Regarding the integrated stellar population properties (Fig.~\ref{fig:intSPP}), we find that red galaxies have very similar distributions regardless of their environment. However, we find that our sample of blue galaxies in groups is biased towards larger values of the mass and older values of the age. This also means redder values of the $(u-r)_\mathrm{int}$ colour, as well as lower values of $\tau/t_0$ indicating shorter star formation episodes. These differences may be caused by the size of the sample, but it is important to take them into account during the analysis of the spatially resolved properties.

\end{document}